\documentclass{elsarticle}

\usepackage{amsmath}  % needed for \tfrac, \bmatrix, etc.
\usepackage{amsfonts}  % needed for bold Greek, Fraktur, and blackboard bold
\usepackage{graphicx}  % needed for figures
\usepackage{epstopdf}  % convert eps figures to pfd (taken care of automatically on modern systems)
\usepackage{amssymb}

\newcommand{\be}{\begin{eqnarray}}
\newcommand{\ee}{\end{eqnarray}}

\newcommand{\myfig}[2]
{\centerline{\resizebox{!}{#1\textwidth}{\includegraphics{#2}}}}

\begin{document}
\ifodd 0
Highlights

Information from matter-wave interferometry and cosmic inflation models is brought together.

The case of scale-invariant gravitational noise leads to a simple tractable result for matter-wave decoherence.

The size of a macroscopic object, whose quantum interference is decohered by
this kind of noise, is very insensitive to the strength and bandwidth of the noise.

Consequently this type of process gives a near-universal limit to one type of macroscopic coherence.

\fi

\title{Matter-wave coherence limit owing to cosmic gravitational wave background}

\author{Andrew M. Steane}
%\email{a.steane@physics.ox.ac.uk} % optional
\ead{a.steane@physics.ox.ac.uk} 
%\affiliation{Department of Atomic and Laser Physics, Clarendon Laboratory, Parks Road, Oxford OX1 3PU, England.}
\address{Department of Atomic and Laser Physics, Clarendon Laboratory, Parks Road, Oxford OX1 3PU, England.}

\date{\today}

\begin{abstract}
We study matter-wave interferometry in the presence of a stochastic background of
gravitational waves.
It is shown that if the background has a scale-invariant spectrum
over a wide bandwidth (which is expected in a class of inflationary models of Big Bang
cosmology), then separated-path interference cannot be observed
for a lump of matter of size above a limit which is very insensitive to
the strength and bandwidth of the fluctuations,
unless the interferometer is servo-controlled or otherwise protected.
For ordinary solid matter this limit is of order 1--10 mm.
A servo-controlled or cross-correlated
device would also exhibit limits to the observation of macroscopic interference,
which we estimate for ordinary matter moving at speeds small compared to $c$.
\end{abstract}

\begin{keyword}
Gravitational wave \sep matter wave \sep interferometry \sep cosmic inflation\sep decoherence
\end{keyword}

%\pacs{03.30.+p, 03.50.De,  04.20.-q, 04.40.Nr}
%PACS numbers: 03.30.+p, 03.50.De,  04.20.-q, 04.40.Nr, 04.40.-b

% 03.30.+p   = Special relativity
% 04.40.-b   = Self-gravitating systems; continuous media and classical fields in curved spacetime

% not 04.25.Nx, 04.30.Db

\maketitle

\section{Introduction}

In a series of papers, Lamine et al. \cite{02Lamine,06Lamine}, have investigated the
effect of a stochastic background
of weak gravitational waves on matter-wave interferometers. We use their
treatment to obtain the following remarkable observation. 
If the early universe  generated gravitational waves with a scale-invariant spectrum
over a wide bandwidth (which is expected in a class of inflationary models of Big Bang
cosmology), we show that there is a near-universal cut-off distance scale for the observation
of matter wave interference. That is, unless the interferometer (as in figure \ref{f.int}a) 
is shielded or otherwise protected from these primordial waves, the fringe visibility will go
rapidly to zero when the radius $R$ of the interfering object
exceeds a value which turns out to be of the order of a
few millimetres for ordinary solid matter. This is not an absolutely
unavoidable
decoherence, only a  technological difficulty, so it is not possible to settle the quantum
measurement problem by this route. However, we also find that 
cross-correlation or servo-control
will not suffice to avoid the decoherene altogether, but merely extend the size or mass 
limit by an amount which depends weakly on the integration time. To avoid this source
of decoherence altogether, it would suffice to shield the interferometer from the
fluctuating gravitational background, but this is very hard to do. 

\section{Calculation}

\begin{figure}
\myfig
%{0.2}
{0.25}
{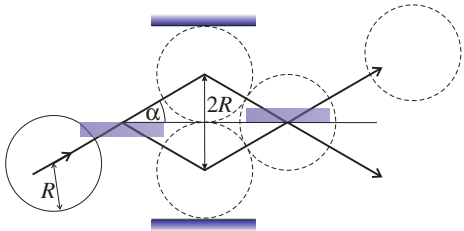}
\caption{A sphere of radius $R$ passes through a two-path interferometer whose
path separation is $2R$. In practice the mirrors and beam-splitters might, for example,
be provided by magnetic or optical forces.}
\label{f.int}
\end{figure}

We are interested in a generic interferometer, but
for the sake of making precise calculations we consider a Mach-Zehnder interferometer in which 
de Broglie waves associated with a sphere of proper radius $R$
interfere, and in which the separation of the arms is $2R$ (see figure \ref{f.int}a). For
such an interferometer, both the interfering object and the separation of the interfering paths
becomes of macroscopic size when $R$ is large enough.

Our treatment follows that of Lamine {\em et  al.} \cite{02Lamine,06Lamine,06Roura}.
In the presence of a stochastic gravitational background with 
a spectral density of strain fluctuations $S_h(\omega)$,
the variance of the phase of a matter-wave interferometer of the type shown in figure \ref{f.int}
is
\be
\Delta \phi^2 = \int_0^\infty \frac{{\rm d}\omega}{2\pi} S_h(\omega) A(\omega) f(\omega)  \label{SAf}
\ee
where $A(\omega)$ is the response of the interferometer to a wave of given frequency, and
$f(\omega)$ is a high pass filter function. For the case where the sphere's group
velocity $v \ll c$, and the wavelength of the gravitational wave is large compared to the
interferometer, $k R \ll 1$, one finds \cite{06Lamine}
\be
A(\omega) =  \left( \frac{4 m v^2}{\hbar\omega} \right)^2 \sin^2(2\alpha) 
\sin^4\left(\frac{\omega\tau}{2}\right)     \label{Afunc}
\ee
where $\tau = R/(v\sin\alpha)$ is the time taken to traverse half of one arm of
the interferometer.

The filter function represents the fact that low frequency `noise'
is not noise but signal---if the fringes move slowly enough then their movement can be tracked by
the interference experiment. We adopt the filter function $f(\omega) = \omega^2/(\omega^2 + \gamma^2)$
where $\gamma$ is the bandwidth. We will discuss this bandwidth after obtaining an
expression for $\Delta\phi^2$ in the presence of stochastic cosmological gravity waves.

\begin{figure}
\myfig
%{0.33}
{0.5}
{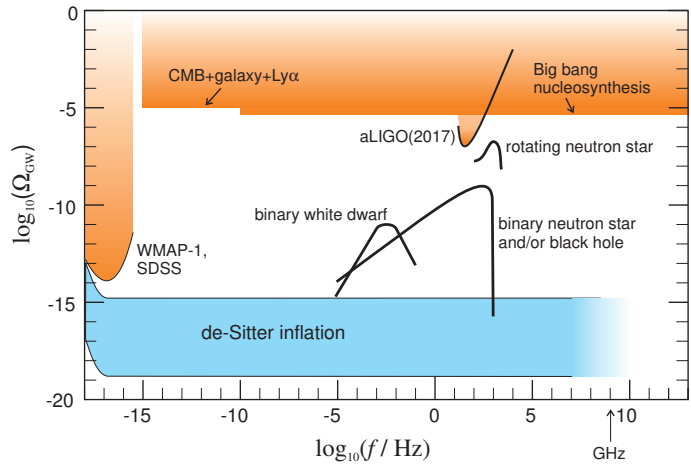}
\caption{Gravitational wave sources. The upper shaded regions are ruled out by various
studies\cite{09LIGO,00Maggiore,06Coward,06Bock,13Hinshaw,06Regimbau,17Abbott,13Thrane}; the lower
shaded region shows, approximately, the range and distribution of primordial cosmological
waves predicted by one type of
inflationary scenario. The curves show the spectra of the main `ordinary' 
sources predicted by standard physics and astronomical surveys.}
\label{f.source}
\end{figure}

Sources of gravitational waves include various `ordinary' processes such as neutron star binaries, black hole binaries,
and rotating neutron stars, and `exotic' processes that are not yet well established but which
are postulated in theoretical descriptions
of early universe physics (figure \ref{f.source})\cite{06Coward,00Maggiore,06Regimbau,06Bock,13Hinshaw}.
We suppose the waves from ordinary sources represent a signal which can
be discriminated by its temporal and other
signatures, and therefore the noise is only given by the exotic sources.
This is outside the present range of confident knowledge, but inflationary models of Big Bang cosmology
suggest that there is now, throughout the universe, a low-level gravitational noise of very
wide bandwidth. Observations of the cosmic microwave background can in principle detect this
gravitational noise at very low frequencies, through the Sachs-Wolfe effect \cite{67Sachs,90Kolb}. This
is not the frequency range we are interested in, but it is useful because it is by
far the most sensitive existing experimental observation. The WMAP and Planck
measurements currently place an upper bound
on the low-frequency gravitational noise close to the level at which inflationary models suggest it
is present, thus it remains undetected but the models remain viable \cite{13Hinshaw}.
%Recently indirect evidence has been reported of very low frequency
%gravitational waves consistent with a primordial origin \cite{BICEP}.

\begin{figure}
\begin{center}
\myfig
%{0.25}
{0.4}
{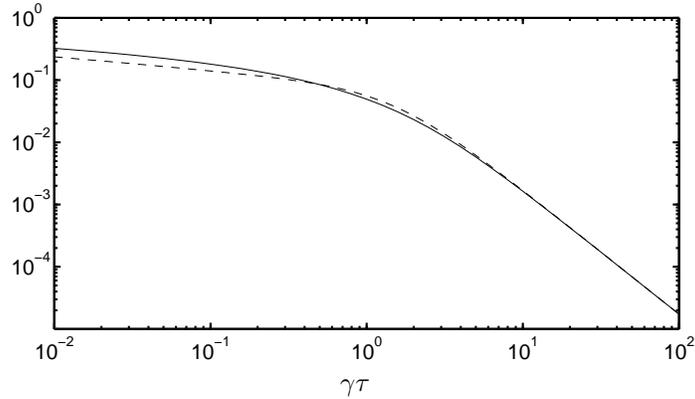}
$\gamma\tau$
\end{center}
\caption{Full curve: integral in Eq. (\protect\ref{Dphiex}) for $\omega_c\tau \gg \pi$; dashed curve:
the function 
%$(\sqrt{3}/10)(1+\gamma^2\tau^2)^{-1}$.
$\sqrt{3}(1+20(\gamma\tau)^{1/4}+ 10\gamma^2\tau^2)^{-1}$.
}
\label{f.integral}
\end{figure}

The cosmological gravitational wave spectrum is usually described in terms
of the measure $\Omega_{\rm GW} = \nu \tilde{\rho}/(\rho_0 c^2) $, where $\rho_0$
is the mass density that would close the universe and $\tilde{\rho}$ is the spectral
energy density of the gravitational radiation (that is, the energy density per unit frequency range
$\Delta \nu$). $\Omega_{\rm GW}$ is dimensionless,
and is related to $S_h(\omega)$ by $S_h = 3 H_0^2 \Omega_{\rm GW}/\omega^3$ where
$H_0 \simeq 2.4 \times 10^{-18}\,{\rm s}^{-1}$ is the Hubble parameter \cite{11Riess}.
The important point for our discussion is
that inflationary models suggest that $\Omega_{\rm GW}$ is {\em independent of $\omega$
over a wide bandwidth}---a property called {\em scale invariance}. The bandwidth is normally
reported in the range $10^7-10^9$ Hz. We will model this by adopting a simple cut-off
at a frequency $\omega_{\rm c}$. 
Hence Eq. (\ref{SAf}) reads
\be
\! \Delta \phi^2 \! &=& \int_0^{\omega_{\rm c}} \frac{{\rm d}\omega}{2\pi} \frac{3 H_0^2 \Omega_{\rm GW}}
{\omega^3 } A(\omega) \frac{\omega^2}{\omega^2 + \gamma^2}  \nonumber  \\
 &=& \frac{24 H_0^2 \Omega_{\rm c} }{\pi \hbar^2} m^2 (v\tau)^4
\int_0^{\omega_{\rm c} \tau} \!\!\!\! \frac{\sin^4 (x/2)}{x^3(x^2 + \gamma^2\tau^2)} {\rm d} x  
\label{Dphiex}
\ee
The integrand in (\ref{Dphiex}) falls quickly once $x > 2\pi$, with the result that the integral is
almost independent of $\omega_{\rm c}$ for $\omega_{\rm c} \tau > 2\pi$. 
It is then well approximated (except at very low $\gamma\tau$) by the expression
%\be
%\Delta \phi^2 \simeq \frac{24 \sqrt{3} H_0^2 \Omega_{\rm GW} }{10 \pi \hbar^2} 
%\frac{ m^2 (v\tau)^4 }{1+ \gamma^2\tau^2}            \label{Dphi}
%\ee
\be
\Delta \phi^2 \simeq \frac{24 \sqrt{3} H_0^2 \Omega_{\rm GW} }{\pi \hbar^2} 
\frac{ m^2 (v\tau)^4 }{1+ 20(\gamma\tau)^{1/4}+10(\gamma\tau)^n}            \label{Dphi}
\ee
where $n=2$ (see figure \ref{f.integral}). The parameter $n$ allows us to consider the effect of 
different filter functions. For example, a perfect high-pass filter with a sharp cut-off at $\omega = \gamma$
results in a response given approximately by Eq. (\ref{Dphi}) with $n=4$.

The above calculation has assumed a point-like model of the interfering sphere, and treats the case
where the gravitational wavelength is large compared to the interferometer, i.e. 
$\omega_{\rm c} R \ll c$ which implies $R \ll 0.05\,$m when 
$\omega_{\rm c} = 2\pi \times 10^9\,{\rm s}^{-1}$. In order to allow for the extended nature of the
interfering sphere, we must consider the fact that, for any given arm of the interferometer, 
different parts of the sphere may experience different amounts of proper time. Consequently the
coordinate associated with the finally measured position of the sphere
may become entangled with internal degrees of freedom, leading
to decoherence when the latter are averaged over \cite{15Pikovski}. A symmetric `bow-tie' design for the
interferometer can cancel the static part of this effect, but for a randomly fluctuating background,
as here, the result is a loss of coherence, and it is one that cannot be avoided by acquiring
data quickly so as to track the fluctuations. Consequently, a reasonable rough model of this is
to extend the validity of Eq. (\ref{Dphi}) to all values of $R$, but insist that the bandwidth $\gamma$
of the interferometer must satisfy $\gamma \ll \pi c/R$, so that high-frequency fluctuations always
contribute to $\Delta \phi^2$. This is a reasonable approach when the arm separation of the interferometer
is similar to the size of the interfering object, as here. 

The result (\ref{Dphi}) has some interesting properties. We have $v\tau=R/\sin\alpha$ so
for any given value of $\gamma\tau$,  (\ref{Dphi}) gives $\Delta\phi \propto m R^2/\sin^2\alpha$---the standard
deviation of the phase is proportional to a quadrupole moment associated with $m$ and the size of
the interferometer. 
The interference fringe visibility falls rapidly to zero once $\Delta\phi  > \pi$.
For a uniform sphere of density $\rho$ one has $m = (4/3)\pi R^3 \rho$, so the critical radius,
above which any interference pattern is washed out, is given by
%\be
%R_c = \left( \frac{15\pi\hbar^2}{256\sqrt{3}H_0^2}\right)^{\!1/10}
%\left(1+\gamma^2\tau^2\right)^{1/10} \rho^{-1/5} \Omega_{\rm GW}^{-1/10}  \label{Rc}
%\ee
\be
R_c = \left( \frac{\sqrt{3}\pi\hbar^2}{512H_0^2}\right)^{\!1/10}
\frac{ \left(1+20(\gamma\tau)^{1/4}+ 10(\gamma\tau)^n \right)^{1/10}}
{ \rho^{1/5} \Omega_{\rm GW}^{1/10} } \label{Rc}
\ee
(taking $\alpha=\pi/4$). As a result of this one tenth power,
$R_c$ is very insensitive to $\Omega_{\rm GW}$.
%Interference will not be
%observable for spheres above a size that depends very little on  $\Omega_{\rm GW}$.

\section{Discussion}

First consider a single interferometer. 
The maximum rate at which data points can be acquired 
is $v/2R = (2\tau \sin \alpha)^{-1}$; this gives a rough estimate of the bandwidth $\gamma$. Hence
for this case $\gamma \tau \lesssim 1$, and in this region $R_c$ is almost independent
of $\gamma\tau$. Hence we can make a simple statement of near-universal validity: there is
a size of a lump of ordinary matter above which interference of the type under discussion
is not observable, unless the inteferometer is somehow shielded from, or servo-controlled
to adjust for, the primordial
cosmological gravity waves (assuming they are there).
We find that for a silica sphere, for example, the critical radius is $3$~mm for
$\Omega_{\rm GW} = 10^{-15}$, and  $8$~mm for $\Omega_{\rm GW}=10^{-19}$
(the sphere's mass is then $m=0.3$ grams and $m=4.6$ grams, respectively).
The values for solid spheres of other common materials are of this same order. 
This rules out an observation of interference of the type under discussion,
unless the interferometer is shielded from such gravity waves, or the local spacetime curvature is
measured rapidly enough to allow the interference pattern to be stabilised by servo-control.

By using many interferometers together, the data rate can be made
arbitrarily high. Then the effective bandwidth is given by the condition that the interferometers must
be spaced by $\ge 2R$ and therefore they are only correlated for fluctuations with 
correlation length $>2R$. 
This implies that the maximum possible bandwidth made available by such cross-correlation
is $\gamma \simeq \pi c/R$, and we have already asserted that in order to allow for the
finite extension of the sphere, the value of $\gamma$ must be taken small compared to this.

We consider the case $\gamma=\pi c/10R$.
Then $\gamma\tau = (\pi/10 \sin\alpha)c/v$, which is $\gg 1$ since the whole discussion has 
assumed $v \ll c$. The result is that the previous values of $R_c$ have to be multiplied by
$(\pi c/10 v)^{n/10}$, so now the critical radius depends on $v$, though not strongly. 
At $v=1\,$m/s one finds $R_c = 0.12\,$m ($4.7\,$m) 
for a silica sphere when $\Omega_{\rm GW}=10^{-15}$ and $n=2$ $(4)$ respectively.
At small $v$ the interferometer is less sensitive to the noise from
gravity waves, but it then becomes more sensitive to other
sources of decoherence such as mechanical instability, black body radiation, magnetic field noise and
collisions with background gas molecules \cite{17Li}. 

\section{Conclusion}

To sum up, the fact that a fluctuating background of gravitational waves will cause decoherence
is not in itself the main result of this paper; both this and 
the mathematical tools to calculate such effects were already established
by previous authors. The main observation of the present work is contained in Eqs (\ref{Dphi})
and (\ref{Rc}). It is that when such a stochastic background has a scale-invariant
spectrum over a wide frequency range, then matter-wave interference 
becomes unobservable, in an unprotected interferometer,
for ordinary solid objects above a size which is independent of their velocity (for $v \ll c$)
and which depends very little on the strength of the stochastic background. This size is of the order
millimetres to centimetres. 

In the case of larger objects, one could in principle still detect interference 
through the use of cross-correlation or servo control methods. However, the present
study suggests that this only extends the size range up to a few metres, for ordinary matter 
traversing the interferometer on a timescale of order seconds.  
We say `suggests' rather than shows since we have only presented a rough estimate of this case.
Nevertheless, these calculations suffice to show that it is 
questionable whether it makes sense to assert that large objects can exhibit
interference effects over large path separations, in the context of the universe as
it is, if indeed the universe has a broad stochastic gravitational wave background. This issue has
already been raised by Lamine {\em et al.} \cite{02Lamine,06Lamine}; our contribution has been to focus attention
on some aspects which are remarkable in their simplicity and generality.

We have used the density of silica (2329 kg$/$m$^3$) for the numerical examples. The critical
distance scale varies as the minus-one-fifth power of this. Very low-density objects of somewhat larger
size may therefore exhibit interference. Applied to a sphere having
the density of the degenerate matter of a neutron star, the calculation gives $R_c \simeq 10\,\mu$m.
It would be interesting to explore the issue, what limits might there be for interference of the matter
waves that are, presumably, associated with a black hole; this is outside the range of applicability
of the methods employed here.

Lamine {\em et al.} \cite{06Lamine} claim that the decoherence associated with this mechanism
constitutes `the quantum--classical transition'
and that `interferometers cannot be shielded' against this type of noise. 
Such claims are not formally justified, because one can in principle propose a thought-experiment
which employed gravitational shielding.
Passive shielding might, for example, employ vast
amounts of viscous matter which absorbed the energy of gravitational waves. Active shielding
might employ a massive distortable membrane plus servo mechanisms, designed to generate
a canceling gravitational wave signal. An interesting avenue to explore is, whether or not there
is any fundamental reason why such schemes could not, in principle, 
succeed in restoring the coherence of
a matter-wave interference experiment. 

Our treatment employed ordinary quantum theory on a classical spacetime described by general relativity.
It therefore makes no proposal regarding new physics beyond that model. However, if a quantum
theory of gravity led to effects that were phenomenologically similar to those of a 
broadband random fluctuation in spacetime, one which was unavoidable, 
then we would have a possible route to settling
the quantum--classical divide. Calculations such as those in this paper and cited works would then
indicate how the resulting limit to interference phenomena relates to the stochastic background.

\bibliography{selfforcerefs,gravitycosmology}
%\bibliography{selfforcerefs}

\end{document}